\documentclass[aps,prd,12pt,preprint,tightenlines,superscriptaddress,
amsfonts,amssymb,amsmath,byrevtex,showpacs,nofootinbib]{revtex4}
\usepackage{graphics}
\usepackage{epsfig,subfigure}
\begin{document}
\newcommand{\dR}{\mathbb R}
\newcommand{\dC}{\mathbb C}
\newcommand{\dS}{\mathbb S}
\newcommand{\dZ}{\mathbb Z}
\newcommand{\id}{\mathbb I}
\newcommand{\dM}{\mathbb M}
\newcommand{\dH}{\mathbb H}
\newcommand{\tm}{\tilde{\mu}}
\newcommand{\tn}{\tilde{\nu}}

\title{Propagation of a string across the cosmological singularity}

\author{Przemys{\l}aw Ma{\l}kiewicz$^\dag$ and W{\l}odzimierz Piechocki$^\ddag$
\\ Department of Theoretical Physics\\So\l tan Institute for Nuclear Studies,
\\ Ho\.{z}a 69, 00-681 Warszawa, Poland;
\\ $^\dag$pmalk@fuw.edu.pl, $^\ddag$piech@fuw.edu.pl}

\date{\today}

\begin{abstract}
Our  results concern  the transition  of a  quantum string through
the singularity of the compactified Milne (CM) space. We restrict
our analysis to the string  winding around the compact dimension
(CD) of spacetime. The CD undergoes contraction to a point
followed by re-expansion. We demonstrate that both classical and
quantum dynamics of considered string are well defined. Most of
presently available calculations strongly suggest that  the
singularity of a time dependent  orbifold is useless as a model of
the cosmological singularity. We believe that our results bring,
to some extent, this claim into question.
\end{abstract}
\maketitle

\section{Introduction}

In the cyclic model of the evolution of the universe the cosmic
singularity (CS) plays the key role because it joins  each two
consecutive classical phases. We propose the following  basic
criterion for the choice of the model of the neighborhood of the
CS: reasonable model should allow for propagation of  quantum
p-brane (i.e., particle, string, membrane,...)   from the
pre-singularity to the post-singularity epoch. If the CS
constitutes an insurmountable obstacle for  quantum elementary
objects, the cyclic evolution cannot be realized.

One of the simplest models of the neighborhood of the CS, inspired
by string/M theory, is the compactified Milne (CM) space. It was
used recently
\cite{Khoury:2001bz,Steinhardt:2001vw,Steinhardt:2001st} in the
cyclic universe scenario.

We have already applied the above criterion for the propagation of
a test particle across the CS of the CM space
\cite{Malkiewicz:2005ii,Malkiewicz:2006wq}. The results we have
obtained suggest that the CM space is a promising model deserving
further investigation by making use of strings and membranes.

In this article we apply our criterion to the evolution of a test
string in the CM space. We examine the propagation of a string in
the so-called winding mode \cite{Pioline:2003bs,Turok:2004gb}. It
is defined to be a state in which a string is winding around the
compact dimension undergoing contraction to a point followed by
re-expansion. It turns out that the propagation is well defined
both at classical and quantum levels.

In Sec. II we recall the classical formalism concerning the
propagation of a test $p$-brane in a fixed background spacetime.
Classical evolution of the string in the winding mode is presented
in Sec. III. Quantization of the dynamics is carried out in Sec.
IV. We conclude in Sec. V.

\section{Classical dynamics of p-brane}

The Polyakov action for a test $p$-brane   embedded in a fixed
background spacetime with  metric $g_{\tm\tn}$ has the form
\begin{equation}\label{act}
    S_p= -\frac{1}{2}\mu_p \int d^{p+1}\sigma
    \sqrt{-\gamma}\;\big(\gamma^{ab}\partial_a X^{\tm} \partial_b
    X^{\tn}
    g_{\tm\tn}-(p-1)\big),
\end{equation}
where $\mu_p$ is a mass per unit $p$-volume,
$(\sigma^a)\equiv(\sigma^0,\sigma^1,\ldots,\sigma^p)$ are
$p$-brane worldvolume coordinates, $\gamma_{ab}$ is the $p$-brane
worldvolume metric, $\gamma := det[\gamma_{ab}]$,
$~(X^{\tm})\equiv (X^\mu, \Theta)\equiv (T,X^k,\Theta)\equiv
(T,X^1,\ldots,X^{d-1},\Theta)$ are the embedding functions of a
$p$-brane, i.e. $X^{\tm} = X^{\tm}(\sigma^0,\ldots,\sigma^p$), in
$d+1$ dimensional background spacetime.

It has been found  \cite{Turok:2004gb} that the total Hamiltonian,
$H_T$, corresponding to the action (\ref{act}) is the following
\begin{equation}\label{ham}
H_T = \int d^p\sigma \mathcal{H}_T,~~~~\mathcal{H}_T := A C + A^i
C_i,~~~~~i=1,\ldots,p
\end{equation}
where $A=A(\sigma^a)$ and $A^i = A^i(\sigma^a)$ are any regular
functions of $p$-volume coordinates,
\begin{equation}\label{conC}
    C:=\Pi_{\tm} \Pi_{\tn} g^{\tm\tn} + \mu_p^2 \;det[\partial_a X^{\tm} \partial_b
    X^{\tn} g_{\tm\tn}]\approx 0,
\end{equation}
\begin{equation}\label{conCi}
    C_i := \partial_i X^{\tm} \Pi_{\tm} \approx 0,
\end{equation}
and where $\Pi_{\tm}$ are the canonical momenta corresponding to
$X^{\tm}$. Equations (\ref{conC}) and (\ref{conCi}) define the
first-class constraints of the system.

The Hamilton equations \cite{Turok:2004gb} are
\begin{equation}\label{hameq}
    \dot{X}^{\tm}\equiv\frac{\partial{X}^{\tm}}{\partial\tau}=
    \{X^{\tm},H_T \},~~~~~~\dot{\Pi}_{\tm}\equiv\frac{\partial{\Pi}_{\tm}}{\partial\tau}=
    \{\Pi_{\tm},H_T \},~~~~~~\tau\equiv\sigma^0,
\end{equation}
where the Poisson bracket is defined by
\begin{equation}\label{pois}
    \{\cdot,\cdot\}:= \int d^p\sigma\Big(\frac{\partial\cdot}{\partial X^{\tm}}
    \frac{\partial\cdot}{\partial \Pi_{\tm}}
     - \frac{\partial\cdot}{\partial \Pi_{\tm}}
    \frac{\partial\cdot}{\partial X^{\tm}}\Big).
\end{equation}

\section{Classical dynamics of  string }

In what follows we restrict our considerations to the compactified
Milne space. Its metric is defined by the line element
\begin{equation}\label{line}
    ds^2 = -dt^2 +dx^k dx_k + t^2 d\theta^2 = \eta_{\mu\nu} dx^\mu
    dx^\nu  + t^2 d\theta^2 = g_{\tm\tn} dx^{\tm} dx^{\tn},
\end{equation}
where $\eta_{\mu\nu}$ is the Minkowski metric, and $\theta$
parameterizes   a circle\footnote{Orbifolding  $\dS^1$ to the
segment $~\dS^1/\dZ_2~$  gives a model of spacetime in the form of
two orbifold planes  which collide and re-emerge at $t=0$. Such
model of spacetime was used in
\cite{Khoury:2001bz,Steinhardt:2001vw,Steinhardt:2001st}, but our
results do not depend on the choice of  topology of the compact
dimension. }.

The spacetime with the metric (\ref{line}) is singular at $t=0$
and it is not a coordinate choice singularity: one of its space
dimensions (denoted by $\theta $) disappears  for a moment at
$t=0$ leading to the degeneracy of the metric. It is a spacelike
orbifold singularity (see \cite{Malkiewicz:2005ii} for more
details).

In  the paper we analyze the dynamics of a string  in the {\it
zero-mode} (the lowest energy state) which is {\it winding} around
the $\theta$-dimension. The string in such a state is defined by
the conditions
\begin{equation}\label{con1}
    \sigma^p := \theta \equiv\Theta~~~~~~\mbox{and}~~~~~
    \partial_\theta X^\mu =0=\partial_\theta
    \Pi_\mu,
\end{equation}
which lead to
\begin{equation}\label{con2}
    \frac{\partial}{\partial\theta}(X^{\tm})=
    (0,\ldots,0,1)~~~~~\mbox{and}~~~~~\frac{\partial}{\partial\tau}(X^{\tm})=
    (\dot{T},\dot{X}^k,0).
\end{equation}

The condition (\ref{con1}) reduces  (\ref{conC})-(\ref{pois}) to
the form in which the canonical pair $(\theta,\Pi_\theta)$ does
not occur \cite{Turok:2004gb}. Thus, a string in the winding mode
is described by (\ref{conC})-(\ref{pois}) with $\tm,\tn,\ldots$
replaced by $\mu,\nu,\ldots$. In fact, the propagation of a string
`reduces' effectively to its  evolution in the Minkowski space
with dimension $d$, while $d+1$ was the original dimension of
spacetime \cite{Turok:2004gb}.

The constraint equations, (\ref{conC}) and (\ref{conCi}), for a
string winding around the $\theta$-dimension reduce to the
following form
\begin{equation}\label{cond1}
    C = \Pi_\mu(\tau)\;\Pi_\nu (\tau)\;\eta^{\mu\nu}
    + \check{\mu}_1^2 \;t^2(\tau)\approx 0,~~~~~~C_1 = 0,
\end{equation}
where $\check{\mu}_1 \equiv 2\pi \mu_1$, and the equations of
motion (\ref{hameq}) read
\begin{equation}\label{eqp}
    \dot{\Pi}_t(\tau)= - 2 A(\tau)\;\check{\mu}_1^2\;T(\tau),~~~~~~
    \dot{\Pi}_k (\tau)= 0,
\end{equation}
and
\begin{equation}\label{eqx}
    \dot{T} (\tau)= - 2 A(\tau)\;\Pi_t(\tau),~~~~~~\dot{X}^k (\tau)=
    2 A(\tau)\;\Pi_k(\tau),
\end{equation}
where $A=A(\tau)$ is any regular function.

It can be verified that in the gauge $A(\tau)=1$, the solutions
are
\begin{equation}\label{solp}
 \Pi_t(\tau)= b_1 \exp (2\check{\mu}_1 \tau) + b_2 \exp (-2\check{\mu}_1
    \tau),~~~~~~\Pi_k(\tau)= \Pi_{0k},
\end{equation}
where $~\;b_1, b_2, \Pi_{0k} \in \dR$, and
\begin{equation}\label{solx}
    T(\tau)= a_1 \exp (2\check{\mu}_1 \tau) + a_2 \exp (-2\check{\mu}_1
    \tau),
    ~~~~~X^k (\tau)= X^k_0 + 2 \Pi_{0k} \;\tau,
\end{equation}
where $\;a_1, a_2, X^k_0 \in \dR$.

To analyze  the propagation of a string across the singularity
$t=0$, we eliminate $\tau$ from (\ref{solp}) and  (\ref{solx}).
Making the choice of $a_1$ and $a_2$ in such a way that $a_1 a_2
<0$ leads to one-to-one relation between $T$ and $\tau$. For
instance, one may  put
\begin{equation}\label{coef}
    a_1 = - a_2 = \sqrt{\Pi_0^k \Pi_{0k}}/2 \check{\mu}_1,
\end{equation}
that gives
\begin{equation}\label{elim}
      T(\tau) =  \sqrt{\Pi_0^k
     \Pi_{0k}}\;\sinh(2\check{\mu}_1\;\tau)/\check{\mu}_1 ,
\end{equation}
which can be rewritten as
\begin{equation}\label{jaw}
    \tau = \frac{1}{2 \check{\mu}_1}\sinh^{-1}\Big( \frac{\check{\mu}_1}
    {\sqrt{\Pi_0^k \Pi_{0k}}}\;t \Big),
\end{equation}
due to $T = t$. The insertion of (\ref{jaw}) into (\ref{solx})
gives
\begin{equation}\label{sols}
 X^k (t)= X^k_0 + \frac{\Pi_0^k}{ \check{\mu}_1}\sinh^{-1}\Big( \frac{\check{\mu}_1}
    {\sqrt{\Pi_0^k \Pi_{0k}}}\;t \Big).
\end{equation}
The solution (\ref{sols}) is bounded and continuous at the
singularity. Thus, the classical dynamics of the zero-mode winding
string is well defined in the CM space.

\section{Quantum dynamics of  string }

In the gauge $A=1$, the Hamiltonian of a string is
\begin{equation}\label{ham1}
    H_T  = C = \Pi_\mu(\tau)\;\Pi_\nu (\tau)\;\eta^{\mu\nu}
    + \check{\mu}_1^2 \;t^2 .
\end{equation}
The quantum Hamiltonian corresponding to (\ref{ham1}) has the form
\cite{Malkiewicz:2006wq}
\begin{equation}\label{qham1}
    \hat{H}_T = \frac{\partial^2}{\partial t^2}- \frac{\partial^2}
    {\partial X^k \partial X_k} + \check{\mu}_1^2 t^2 ,
\end{equation}
owing to  $T=t$.

According to the Dirac quantization method \cite{PAM,MHT} the
physical states $\psi$ should first of all satisfy, due to
(\ref{cond1}) and (\ref{ham1}), the equation
\begin{equation}\label{eqs}
   \hat{H}_T \;\psi (t,X^k) = 0.
\end{equation}
To solve (\ref{eqs}),  we make the substitution
\begin{equation}\label{sub}
    \psi(t,X^1,\ldots,X^{d-1})= F(t)\;G_1(X^1)\;G_2(X^2)\cdots
    G_{d-1}(X^{d-1}),
\end{equation}
which turns (\ref{eqs}) into the following set of equations

\begin{equation}\label{eqG}
   \frac{d^2 G_k (q_k,X_k)}{dX_k^2}+ q_k^2 \;G_k(q_k,X_k) = 0,~~~~~~~k=1,
   \ldots,d-1 ,
\end{equation}
\begin{equation}\label{eqF}
    \frac{d^2 F(q,t)}{dt^2}+(\check{\mu}_1^2 t^2 + q^2)\;F(q,t)
    =0,~~~~~~q^2 := q_1^2 +\ldots +q_{d-1}^2,
\end{equation}
where $q_k^2, q^2 \in \dR$  are the separation constants.

Two independent  solutions to (\ref{eqG}) have the form
\begin{equation}\label{solG}
G_{1k}(q_k,X_k)= \cos (q_k X^k),~~~~~G_{2k}(q_k,X_k)= \sin (q_k
X^k),~~~~~~~~k=1,\ldots,d-1
\end{equation}
(there is no summation in $~q_k X^k~$ with respect to $k$).

Two  independent  solutions of (\ref{eqF}) read \cite{SWM}
\begin{equation}\label{solF1}
   F_1 (q,t)= \exp{(-i \check{\mu}_1 t^2/2)}\;H \Big(-\frac{\check{\mu}_1+iq^2 }
   {2 \check{\mu}_1}, (-1)^{1/4}\;\sqrt{\check{\mu}_1} \;t \Big),
\end{equation}
\begin{equation}\label{solF2}
 F_2 (q,t)=  \exp{(-i \check{\mu}_1 t^2/2)}\;_1F_1
 \Big(\frac{\check{\mu}_1+iq^2 }{4 \check{\mu}_1},\frac{1}{2},i
\check{\mu}_1 t^2 \Big),
\end{equation}
where $H(a,t)$ is the Hermite function and $_1F_1(a,b,t)$ denotes
the Kummer confluent hypergeometric function.

In what follows we present the construction of a Hilbert space,
$\mathcal{H}$,  of our system based on the solutions
(\ref{solG})-(\ref{solF2}):

\noindent First, we intend to redefine (\ref{solF1}) and
(\ref{solF2}) to get {\it bounded} functions on $\dR \times
[-t_0,t_0]$, where $[-t_0,t_0]$ denotes the `time-like'
neighborhood of the singularity.  For fixed value of $q$ and
$t\in[-t_0,t_0]$ the solutions (\ref{solF1}) and (\ref{solF2}) are
bounded functions, as it is demonstrated by the plots of  Fig.
(1).

\begin{figure}[h]
\centering \subfigure {\includegraphics[width=2.3in]{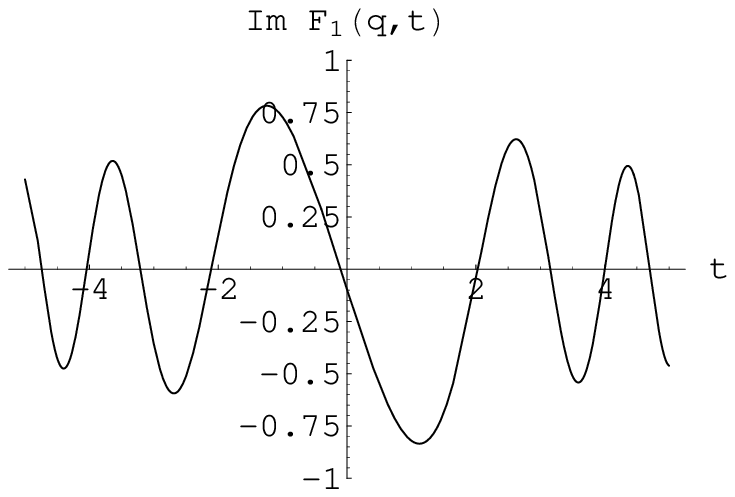}}
\hspace{0.4in} \subfigure
{\includegraphics[width=2.3in]{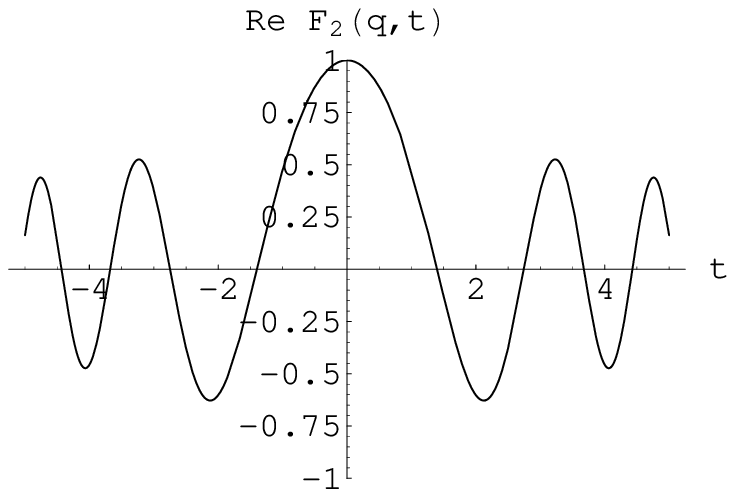}} \caption{Solutions as
functions of $t$ in the neighborhood of the singularity
($\check{\mu}_1=1,$  $q=1$).}
\end{figure}

For $q^2\gg \check{\mu}_1^2 t_0^2$, the solution to (\ref{eqF})
can be approximated by

\begin{equation}\label{solFbigq}
F(q,t)\approx A(q)\sin(qt)+B(q)\cos(qt),
\end{equation}
where $A(q)$ and $B(q)$ are any functions. Finding bounded $A(q)$
and $B(q)$ in (\ref{solFbigq}) gives bounded $F(q,t)$.  They can
be determined from  the equations ($q^2\gg \check{\mu}^2t_0^2$)
\begin{equation}
F(q,t)|_{t=0}=B(q)~~~~\mbox{and}~~~~\partial_tF(q,t)_{|_{t=0}}=qA(q).
\end{equation}
It can be checked \cite{SWM} that
\begin{equation}\label{re}
\begin{array}{ll}
F_1(q,t)_{|_{t=0}}=\frac{\sqrt{\pi}~2^{\frac{-\imath
q^2-\check{\mu}_1}{2\check{\mu}_1}}}
{\Gamma(\frac{3}{4}+\imath\frac{q^2}{4\check{\mu}_1})}, &~~~~
\partial_tF_1(q,t)_{|_{t=0}}=\frac{(-1)^{-1/4}\sqrt{\pi}~(-\imath q^2-\check{\mu}_1)
~2^{\frac{-\imath
q^2-\check{\mu}_1}{2\check{\mu}_1}}}{2\sqrt{\check{\mu}_1}~\Gamma(\frac{5}{4}+
\imath\frac{q^2}{4\check{\mu}_1})},\\
F_2(q,t)_{|_{t=0}}= 1,&~~~~\partial_tF_2(q,t)_{|_{t=0}}=0.
\end{array}
\end{equation}
It results from   (\ref{re}) that the solution $F_2(q,t)$ is a
bounded function, so it does not need any redefinition. For $q^2$
big enough, $F_1(q,t)_{|_{t=0}}$ and
$\partial_tF_1(q,t)_{|_{t=0}}$ are found to be (see Eq. (6.1.45)
in \cite{abram})
\begin{equation}
|F_1(q,t)_{|_{t=0}}|\approx
\sqrt[4]{\frac{\check{\mu}_1}{4}}~\frac{\exp{(\frac{\pi}{8\check{\mu}_1}q^2)}}
{\sqrt{q}},~~~~|\partial_tF_1(q,t)_{|_{t=0}}|\approx
\sqrt[4]{\frac{\check{\mu}_1}{4}}~\sqrt{q}\exp{(\frac{\pi}{8\check{\mu}_1}q^2)}.
\end{equation}
Thus, we redefine the  solution $F_1(q,t)$ as follows
\begin{equation}\label{redef}
F_1 (q,t):=
\sqrt{q}\;\exp{(-\frac{\pi}{8\check{\mu}_1}q^2)}\exp{(-i
\check{\mu}_1 t^2/2)}\;H \Big( -\frac{\check{\mu}_1+iq^2 }
   {2 \check{\mu}_1}, (-1)^{1/4}\;\sqrt{\check{\mu}_1} \;t\Big).
\end{equation}
It is clear that  (\ref{redef}) is the solution of (\ref{eqF})
owing to the structure of the equation. Now, one can verify that
\begin{equation}
\begin{array}{ll}
|A_1(q)|=\sqrt[4]{\frac{\check{\mu}_1}{4}},&~~~~|B_1(q)|=\sqrt[4]{\frac{\check{\mu}_1}{4}},
\\A_2(q)=0,&~~~~B_2(q)=1.
\end{array}
\end{equation}
Therefore, we get the result that the functions $\dR\times
[-t_0,t_0] \ni(q,t)\rightarrow F_s (q,t)\in \dC,~~(s=1,2)~$ are
bounded.

\noindent Second, we define the following generalized functions
\begin{equation}\label{hs}
h_s(t,X^1,\ldots,X^{d-1}):=
\int_{\dR^{d-1}}f(q_1,\ldots,q_{d-1})\;F_s(q,t)\prod_k \exp(-i q_k
X^k)\;dq_1\ldots
   dq_{d-1},
\end{equation}
where $~q^2 = q_1^2 + \dots q_{d-1}^2,~$ and where $f\in
L^2(\dR^{d-1})~$. Since $~F_s~$ are bounded, the functions $f F_s
\in L^2([-t_0,t_0]\times \dR^{d-1})$.  Equation (\ref{hs})
includes (\ref{solG}) due to the term $~\exp(-i q_k X^k)$, with
$q_k \in \dR$.

\noindent Finally, we notice that (\ref{hs}) defines the Fourier
transform of $~fF_s$. Therefore, according to the Fourier
transform theory  (see, e.q. \cite{LDP}) the equation (\ref{hs})
defines the mapping
\begin{equation}\label{plan}
 L^2(\dR^{d-1})\ni f \longrightarrow h_s \in  L^2([-t_0,t_0]
 \times \dR^{d-1})=: \tilde{\mathcal{H}}.
\end{equation}
Replacing  $f$ by consecutive elements of an orthonormal  basis in
$L^2(\dR^{d-1})$ leads to an infinite countable  set of vectors in
$\tilde{\mathcal{H}}$. So obtained set of vectors can be
rearranged into a set of independent vectors and further turned
into an orthonormal basis  by making use of the Gram-Schmidt
procedure \cite{EP}. One can show \cite{LDP} that the span of such
an orthonormal basis, $\mathcal{F}$, is dense in
$\tilde{\mathcal{H}}$. The completion of $\mathcal{F}$ defines the
Hilbert space $\mathcal{H}\subseteq \tilde{\mathcal{H}}$.

To illustrate the above construction, let us use the Hilbert space
$L^2(\dR^{d-1}):= \bigotimes_{k=1}^{d-1}L^2_k(\dR)$, where
$L^2_1(\dR)= L^2_2(\dR)= \ldots = L^2_{d-1}(\dR)\equiv L^2(\dR)$.
Let us take a countable infinite set of vectors $f_n \in L^2(\dR)$
defined as
\begin{equation}\label{basis}
    f_n(q):= \frac{1}{\sqrt{2^n n! \sqrt{\pi}}}\;\exp(-q^2/2)\;H_n(q),
    ~~~~~n=0,1,2,\ldots,
\end{equation}
where $H_n(q)$ is the Hermite polynomial. It is proved in
\cite{NIA} that (\ref{basis}) defines an orthonormal basis in
$L^2(\dR)$. The basis (\ref{basis}) can be used to construct a
basis in  $L^2(\dR^{d-1})$. The basis is defined as the set of all
vectors of the form $\bigotimes_{k=1}^{d-1} f_{n_k}(q^k)\in
L^2(\dR^{d-1})$.  Further steps of the procedure leading to the
dense subspace $\mathcal{F}$ are the same as described in the
paragraph including Eq. (\ref{plan}).

It is clear that (\ref{hs}), owing to the above construction,
defines the solution to the equation $\hat{H}_T h_s = 0$. It is
obvious that the Hamiltonian (\ref{qham1}) is bounded from below
(and above) and self-adjoint on the Hilbert space $\mathcal{H}$.

\section{Conclusions}

Our results  demonstrate that the dynamics of a {\it test}
string\footnote{A test string, by definition, does not change the
background spacetime.} winding around the compactified dimension
in the {\it zero-mode} state is well defined. At the classical
level there are no instabilities near the singularity
(characteristic for particle dynamics) and our non-perturbative
method of quantization leads to a non-singular quantum evolution
of a string\footnote{ For the discussion of some physical aspects
of propagation of winding states through the Milne space
singularity we recommend \cite{Russo:2003ky}. One shows that the
simple string cosmology based on the Milne space can be used to
obtain interesting cosmologies in four dimensions.}.

A great challenge is  examination of the dynamics of a {\it
physical} string, i.e. a string that may change its state and
modify the background spacetime during the evolution of the entire
system. The issue of the backreaction has been intensively studied
recently (see \cite{Craps:2006yb} for review). Calculations of the
S-matrix amplitudes in  time dependent singular orbifolds are
plagued by infinities (see, e.g.
\cite{Liu:2002kb,Lawrence:2002aj,Berkooz:2002je}). However, there
exist some attempts to cure the divergencies of scattering
amplitudes (see, e.g. \cite{Cornalba:2003kd}). It seems to us that
such results only mean that in these type of orbifolds the string
{\it perturbation} theory breaks down and one should apply {\it
non-perturbative} methods for analyses.

It is claimed in \cite{Horowitz:2002mw} that the pessimistic
results concerning the S-matrix can be confirmed by
non-perturbative analyses.  One shows in \cite{Horowitz:2002mw}
that, for instance, a single particle added to a time dependent
singular orbifold causes the orbifold to collapse into a large
black hole, which leads finally to the creation of a big-crunch
that would not be followed by a big-bang\footnote{ See
\cite{Nekrasov:2002kf} for related considerations, where analyses
of the bosonic and fermionic spectrum in the Milne orbifold lead
the author to the conclusion that the resolution of the
cosmological orbifold singularity is impossible.}. This result
suggests that the compactified Milne space is useless as a model
of the neighbourhood of the cosmological singularity in the
context of the cyclic universe scenario.

It seems to us that the result of \cite{Horowitz:2002mw} should be
treated with caution. It is so because it was obtained by making
use of general relativity, i.e. a classical theory, which  is not
suitable for {\it complete} understanding of the physics of a
black hole.

In our article we have proposed a new non-perturbative method of
analysis of the singularity of a time dependent orbifold. We
believe that it can be further applied to analyses of the
propagation of a winding string in {\it non-zero} modes (excited
states). It may initiate  a new direction in a systematic
examination of the backreaction phenomena in the context of time
dependent orbifolds.

We are conscious that {\it real} understanding of the problem
would require quantization of the entire system, including string
and  orbifold, but it seems to be beyond the scope of the current
stage of quantum gravity.

\begin{acknowledgments}
PM would like to thank the European Network of Theoretical
Astroparticle Physics ILIAS/N6 under contract number
RII3-CT-2004-506222 for partial financial support. WP is grateful
to Ben Craps for helpful discussion. We would like to thank  the
anonymous referee for the suggestion of including
\cite{Russo:2003ky,Nekrasov:2002kf} into our list of references.
\end{acknowledgments}

\end{document}